\documentclass[12pt]{article}
\usepackage{epsfig}
\usepackage{mathrsfs}
\usepackage{amsfonts}
\usepackage{amssymb}
\usepackage{graphicx}
\usepackage{amsmath}

\def\beq{\begin{equation}}
\def\eeq#1{\label{#1}\end{equation}}
\def\eeqn{\end{equation}}

\def\beqa{\begin{eqnarray}}
\def\eeqa#1{\label{#1}\end{eqnarray}}
\def\eeqan{\end{eqnarray}}

\let\bar=\overbar

\def\Dslash{\not{\hbox{\kern-4pt $D$}}}
\def\dslash{\not{\hbox{\kern-2pt $\del$}}}

\def\msb{{\bar{\ssstyle M \kern -1pt S}}}

\usepackage{fancyhdr,graphicx}
\def\Title#1{\begin{center} {\Large {\bf #1} } \end{center}}

\begin{document}

\Title{Clustered Quark Matter Calculation for Strange Quark Matter}

\bigskip\bigskip


\begin{raggedright}

{\it Xuesen Na\index{Na, X.S.}\\
Department of Astronomy\\
School of Physics\\
Peking University\\
Beijing 100871\\
P. R. China\\
{\tt Email: naxuesen@pku.edu}}
\bigskip\bigskip
\end{raggedright}

\abstract Motivated by the need for a solid state strange quark
matter to better explain some observational phenomena, we discussed
possibility of color singlet cluster formation in cold strange quark
matter by a rough calculation following the excluded volume method
proposed by Clark et al (1986) and adopted quark mass density
dependent model with cubic scaling. It is found that 70\% to 75\% of
volume and 80\% to 90\% of baryon number is in clusters at
temperature from 10MeV to 50MeV and 1 to 10 times nuclear density.

\section{Introduction}

Many believe that neutron stars might be hybrid star with quark
matter at its core or even composed entirely of quark matter with 2
or 3 flavor quark matter. However due to non-perturbative nature of
QCD, property of quark matter at relevant temperature and baryon
number density for neutron star is still far from clear. Even though
at high enough density when asymptotic freedom sets in, quark matter
should appear in color superconducting phase (CFL or 2SC) as is
calculate rigorously from first principles \cite{alford08}, no one
can be sure about to what degree this sector can stretch toward
lower density on the QCD phase diagram. On astrophysics side, there
are some hints such as the need for large amount of energy during
bursts of SGRs and possible precession signals of some radio pulsars
suggesting a solid phase of pulsar interior~\cite{xu09}. Therefore,
in order to have a quark matter phase with regular lattice like
normal solid seen on earth it is intriguing to discuss the
possibility of quark clustering at moderate densities where quark
clusters can serve as lattice points just like positive ions in
metal.

\section{A Simple Calculation}

\begin{figure}[htb]
\begin{center}
\epsfig{file=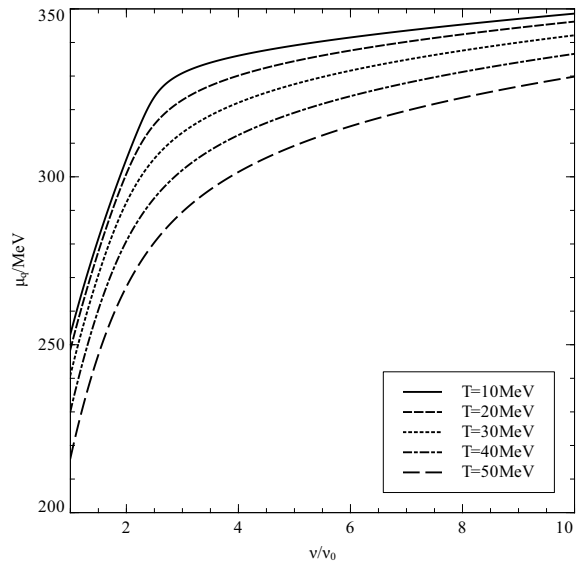,height=3in} \caption{Chemical potential with
fixed cluster mass $M_{cl}=1000$MeV at $1\sim 10\nu_0$ and $T=10\sim 50$MeV}
\label{fig:naivemuq}
\end{center}
\end{figure}

\begin{figure}[htb]
\begin{center}
\epsfig{file=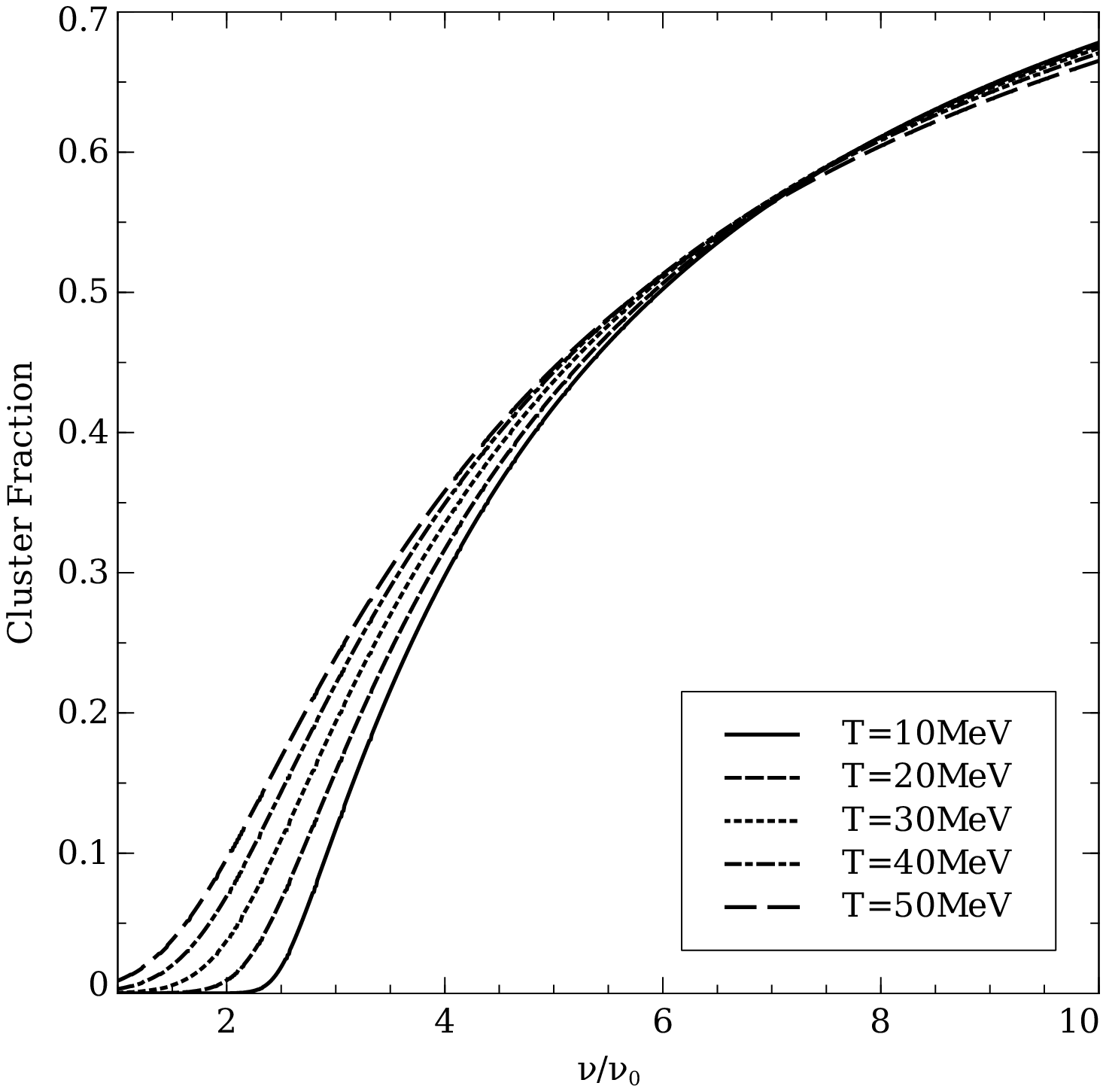,height=3in} \caption{Cluster fraction with
fixed cluster mass $M_{cl}=1000$MeV at $1\sim 10\nu_0$ and $T=10\sim 50$MeV}
\label{fig:naiveclusterfraction}
\end{center}
\end{figure}

We first perform some simple calculation to incorporate clustering in three flavor quark
matter modeled by non-interacting relativistic Fermi gas. Consider color-singlet spin-$1/2$
clusters with 3 quarks since quarks interact strongly attractively in this channel. As a first
approximation we can fix the cluster mass at $M_{cl}=1000$MeV which is roughly the average mass of baryon octet.
Then we use Quark Mass Density Dependent (QMDD) model with parametrization following Peng et al.~\cite{peng99}
\begin{align}\label{eq:QMDD}
M_u&=M_d=\frac{D_0}{\nu^{1/3}}\\
M_s&=m_{s0}+\frac{D_0}{\nu^{1/3}}
\end{align}
to simulate asymptotic freedom and confinement. Suppose clusters can
dissolve into free quarks in the Fermi sea and vice versa, we have chemical equilibrium relation
\begin{equation}\label{eq:chemeq}
\mu_c=3\mu_q
\end{equation}
and treating clusters as a new ingredient of the Fermi sea, we now
have
\begin{equation}\label{eq:nu1}
\nu=\int\frac{d^3p}{(2\pi)^3}\sum_{i=c,u,d,s}\frac{g_i}{\exp\left(\frac{\sqrt{p^2+M_i^2}-\mu_i}{T}\right)+1}
\end{equation}
where the degeneracy factors are $g_u=g_d=g_s=6$ and $g_c=16$ for
all settings of in the baryon octet. Now we can solve equation
(\ref{eq:QMDD}), (\ref{eq:chemeq}) and (\ref{eq:nu1}) to get $\mu_q$
for fixed temperature $T$ and baryon number density $\nu$. Fig
\ref{fig:naivemuq},\ref{fig:naiveclusterfraction} shows how chemical
potential $\mu_q$ and cluster fraction which is the fraction of
baryon number in clusters as a function of baryon number density
$\nu$ in unit of nuclear density $\nu_0=0.159 fm^{-1}$ at relatively
low temperature $T=10\sim 50$MeV (Here we extend the temperature up
to $50$MeV to show the effect of temperature although for neutron
stars after a few second old we should have a temperature lower than
$10$MeV). The parameters in QMDD model equation (\ref{eq:QMDD}) were
adopted from~\cite{peng99} to be $D_0=(80\mbox{MeV})^2$ and
$m_{s0}=150$MeV. As we can see $\mu_q$ first rise rapidly before
reaching about $300\mbox{MeV}$, while cluster fraction rise rapidly
after $\mu_q$ has reached $300\mbox{MeV}$ then gradually saturate.

\begin{figure}[htb]
\begin{center}
\epsfig{file=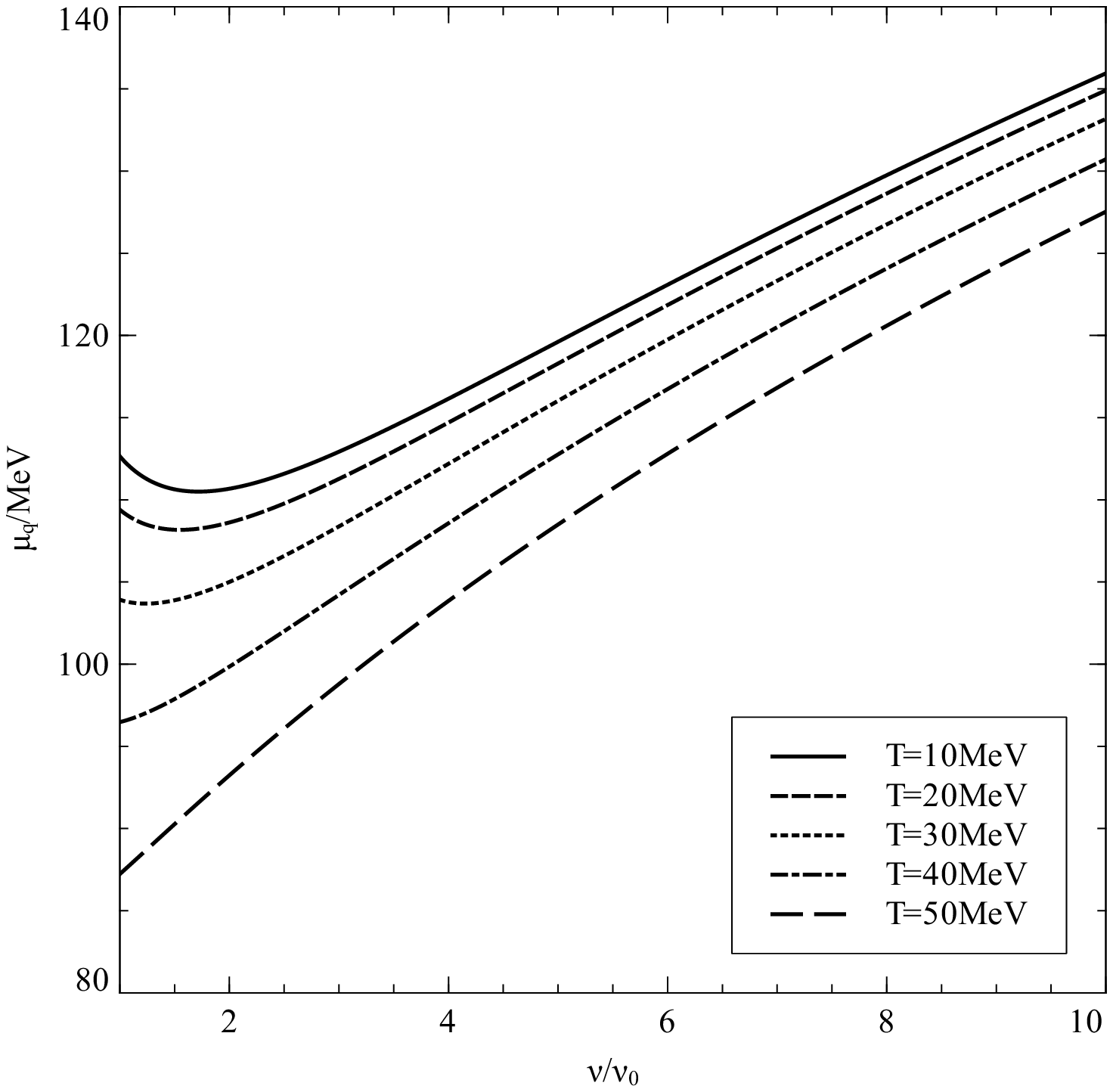,height=3in} \caption{Chemical potential with
cluster mass $M_{cl}=2M_u+M_s$ at $1\sim 10\nu_0$ and $T=10\sim 50$MeV}
\label{fig:McDDmuq}
\end{center}
\end{figure}

\begin{figure}[htb]
\begin{center}
\epsfig{file=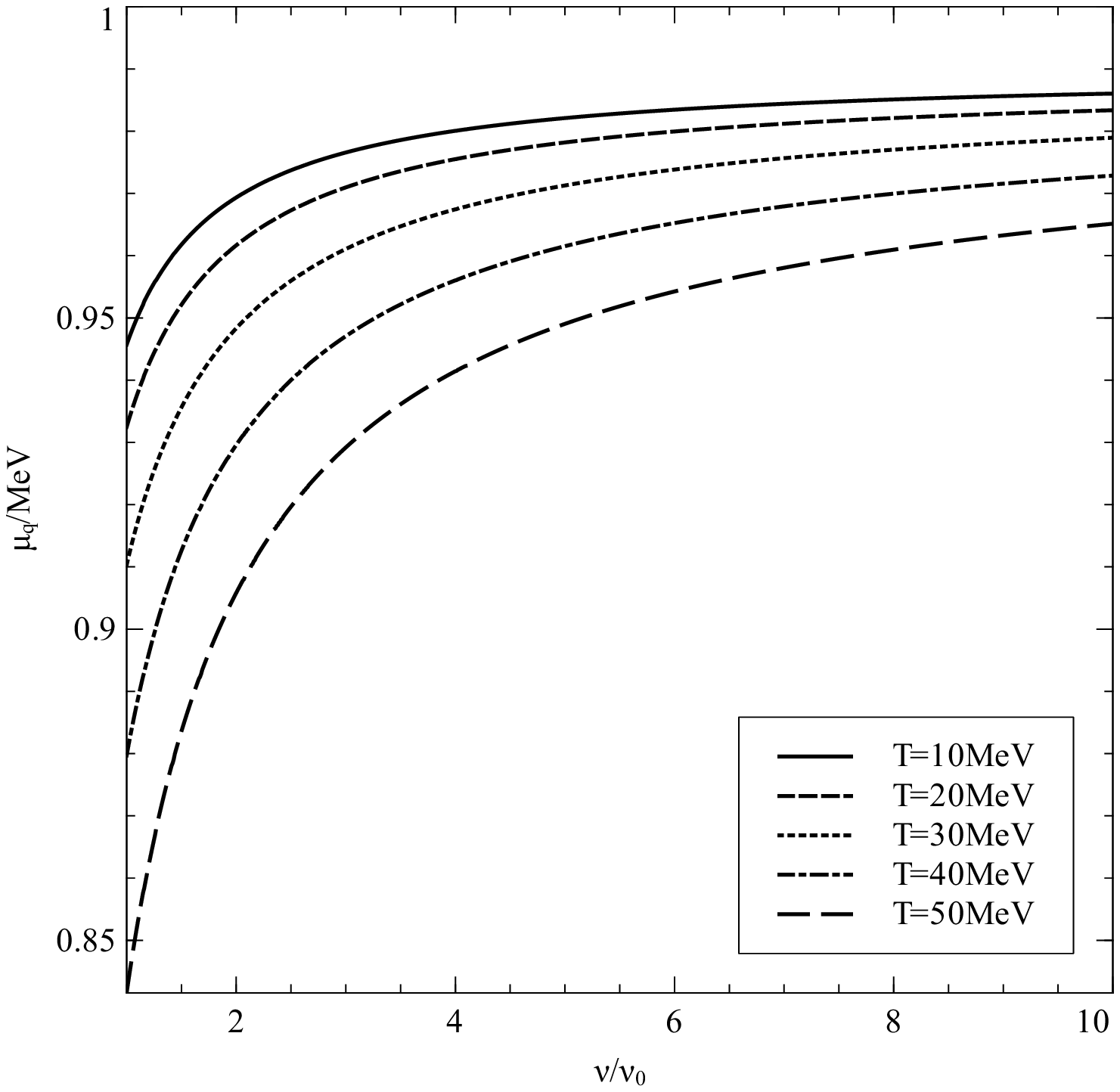,height=3in} \caption{Cluster fraction with
cluster mass $M_{cl}=2M_u+M_s$ at $1\sim 10\nu_0$ and $T=10\sim 50$MeV}
\label{fig:McDDclusterfraction}
\end{center}
\end{figure}

The above simple calculation have two shortcomings
\begin{enumerate}
\item Cluster mass is fixed, while it is expected to rise with
increasing density since quark mass would gradually grow in QMDD
model.
\item Interaction between clusters and quarks is not taken into
account
\end{enumerate}
To introduce the density dependence of cluster mass
we can let the cluster mass be the sum of mass of
two u or d quarks and an s quark:
\begin{equation}
M_{cl}(\nu)=2M_u(\nu)+M_s(\nu)
\end{equation}The resulting chemical potential and cluster fraction
in total baryon number are shown in Fig \ref{fig:McDDmuq} and
\ref{fig:McDDclusterfraction} respectively. As we can see from the plot,
the chemical potential is significantly lowered because of
much lower cluster mass as sum of density dependent quark masses (since
QMDD quark mass in this density range is must less than $\sim300$MeV),
thus from relatively low densities cluster fraction is already very high.

\section{Excluded Volume Method}
One way to remedy the second shortcoming is to introduce an excluded
volume method i.e. consider influence of cluster's finite volume on
momentum space integral. Similar method was first used by Clark et al~\cite{clark86}
at zero temperature and latter by Bi and Shi~\cite{bs87} for finite
temperature partly attempting to explanation to the EMC effect.
Here we do not introduce running coupling constant to account for
interaction among deconfined quarks, but use QMDD.

\begin{figure}[htb]
\begin{center}
\epsfig{file=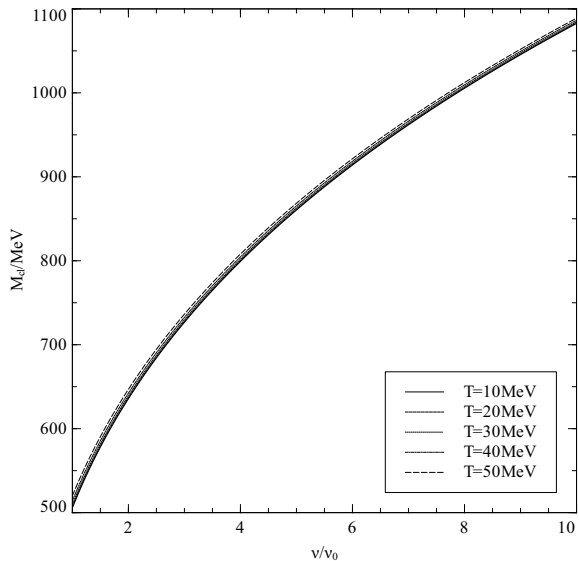,height=3in} \caption{Cluster mass $M_{cl}$ in
excluded volume method at $1\sim 10\nu_0$ and $T=10\sim 50$MeV}
\label{fig:Mc}
\end{center}
\end{figure}

\begin{figure}[htb]
\begin{center}
\epsfig{file=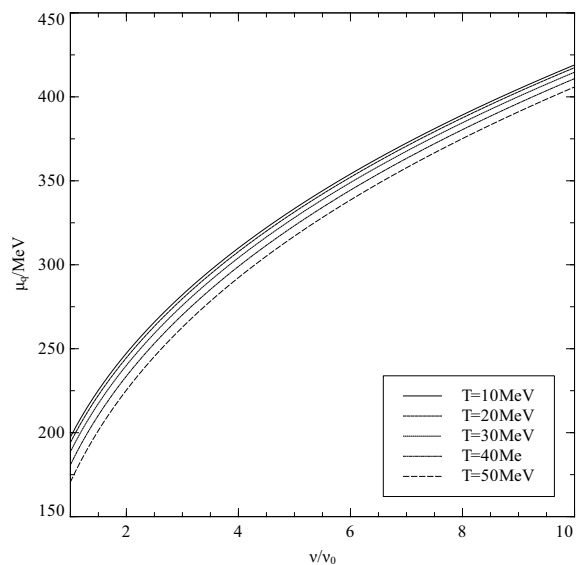,height=3in} \caption{Chemical potential $\mu_q$ in
excluded volume method at $1\sim 10\nu_0$ and $T=10\sim 50$MeV}
\label{fig:muq}
\end{center}
\end{figure}

\begin{figure}[htb]
\begin{center}
\epsfig{file=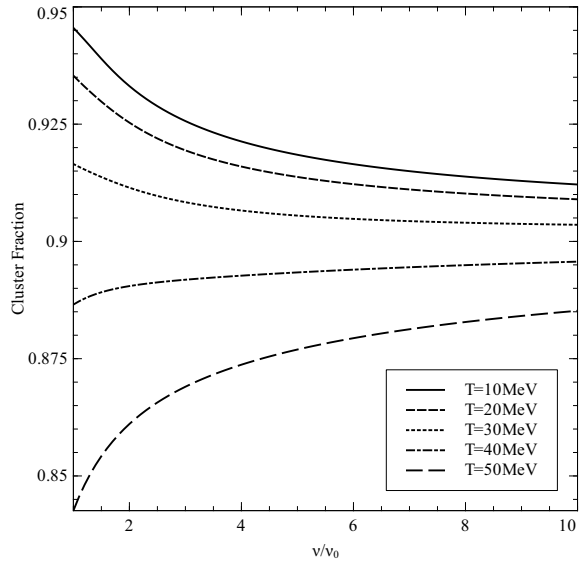,height=3in} \caption{Cluster fraction in
excluded volume method at $1\sim 10\nu_0$ and $T=10\sim 50$MeV}
\label{fig:clusterfraction}
\end{center}
\end{figure}

\begin{figure}[htb]
\begin{center}
\epsfig{file=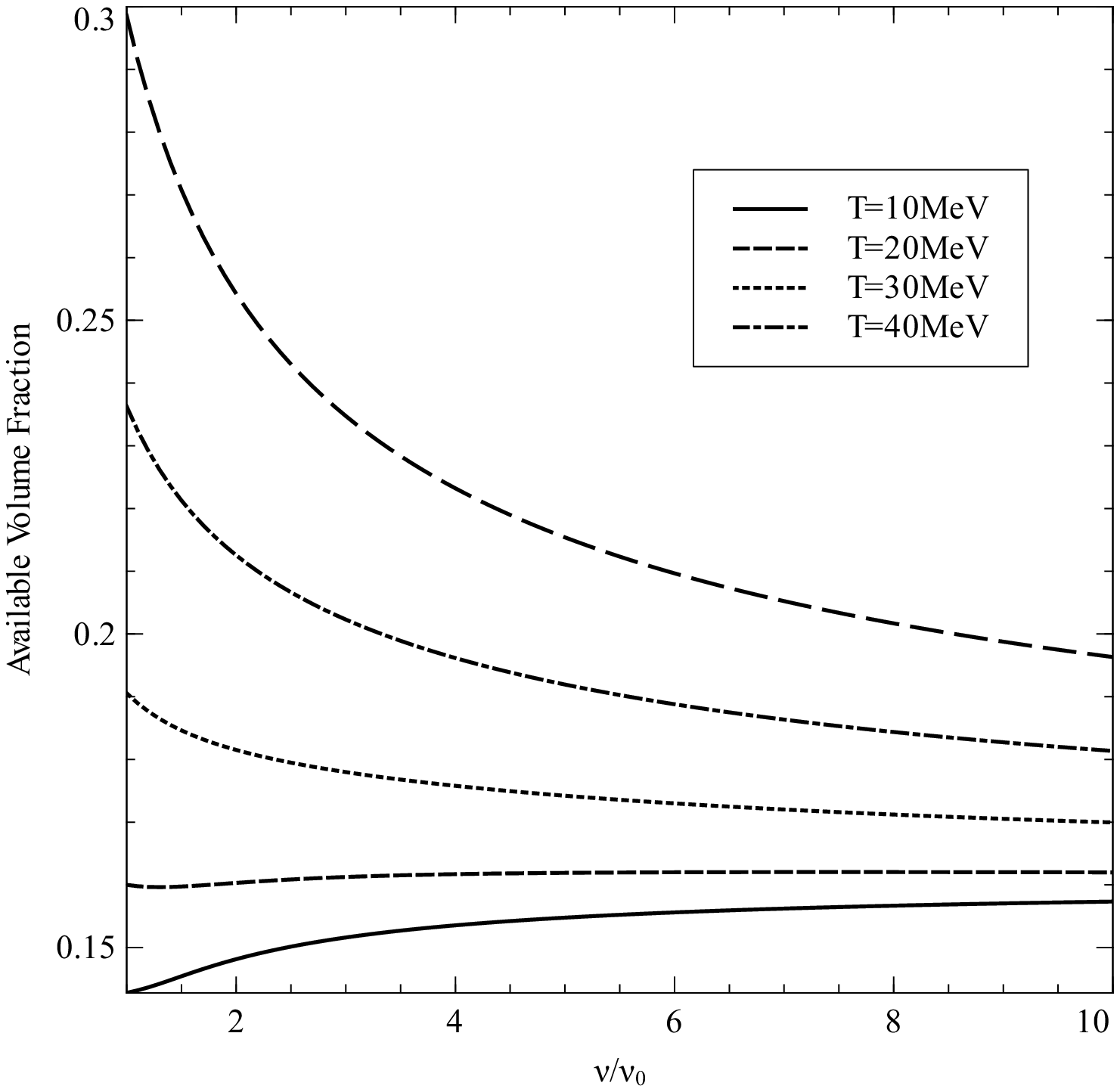,height=3in} \caption{Available volume fraction in
excluded volume method at $1\sim 10\nu_0$ and $T=10\sim 50$MeV}
\label{fig:availablevol}
\end{center}
\end{figure}

Still consider color singlet spin-1/2 3-quark clusters, interaction
between free quarks, between clusters and quarks and among clusters
come in similar to hard-ball potential: as a factor of available
volume multiplied to every momentum integral.
\begin{equation}
\eta=1-\frac{N_{cl}V_{cl}}{V}
\end{equation}
To self-consistently solve cluster mass we will need a relation
between volume and mass. Use the MIT bag model with a simplified
form of total energy just as in~\cite{clark86}
\begin{equation}
M(R)=BV+\frac{c}{R}
\end{equation}
and adopting Bag constant $B=161.5\mbox{MeV}\cdot
fm^{-3}=(187.7\mbox{MeV})^4$ from Saito \& Thomas~\cite{saito95}
from a fit to baryon octet we can get $c$ by requiring $M(R)$ to
have minimum value\footnote{actually the dependence for hadrons with
strangeness is not $c/R$ (T. DeGrand et al 1975~\cite{degrand75})
but the difference is small. Thus we continue to use parametrization
in the massless case.} of $M_\Lambda=1115$MeV as roughly the average
mass in baryon octet.
\begin{equation}
c=\left(\frac{3}{4}\frac{M_\Lambda}{(4\pi
B)^{1/4}}\right)^{4/3}\approx 3.15
\end{equation}
Since hadronic vacuum has been moved out of the entire region of
quark matter, compared to the environment, the cluster (as a MIT
bag) would have an energy $M_{cl}(R)=c/R$ which gives a $M-V$ relation
\begin{equation}
V_{cl}=\frac{4\pi}{3}\left(\frac{c}{M_{cl}}\right)^3
\end{equation}
In addition, the pressure of cluster (which is also pressure of
quark matter) is
\begin{equation}\label{eq:pressure}
P=-\frac{\partial M_{cl}}{\partial V_{cl}}=\frac{M_{cl}^4}{4\pi c^3}
\end{equation}
We also assume the relativistic equation of state following~\cite{clark86}:
\begin{equation}\label{eq:EREoS}
P=\frac{1}{3}\epsilon
\end{equation}where the energy density $\epsilon$ do not include
vacuum energy $B$. Baryon number density, energy density and
available volume factor $\eta$ can be written as
\begin{align}\label{eq:nuepsiloneta}
\nu&=\frac{\eta}{(2\pi)^3}\sum_{i=c,u,d,s}\int dp\frac{4\pi p^2g_i
B_i}{\exp\left(\frac{\sqrt{p^2+M_i^2}-\mu_i}{T}\right)}\\
\epsilon&=\frac{\eta}{(2\pi)^3}\sum_{i=c,u,d,s}\int dp\frac{4\pi
p^2g_i
\sqrt{p^2+M_i^2}}{\exp\left(\frac{\sqrt{p^2+M_i^2}-\mu_i}{T}\right)}\\
\eta&=1-\frac{4\pi}{3}\left(\frac{c}{M_{cl}}\right)^3\int dp \frac{4\pi
p^2g_c}{\exp\left(\frac{\sqrt{p^2+M_{cl}^2}-\mu_c}{T}\right)+1}
\end{align}
Then chemical potential $\mu_q$, cluster mass $M_{cl}$ and radius $R_{cl}$
can be solved from equation (\ref{eq:pressure}), (\ref{eq:EREoS}) and
(\ref{eq:nuepsiloneta}). Fig \ref{fig:Mc},\ref{fig:muq}, \ref{fig:clusterfraction} and
\ref{fig:availablevol}
shows various quantities with temperature $T=10\sim 50$MeV as a function of
baryon number density in the range $1\sim 10\nu_0$.

\section{Results and Discussion}
For baryon number density in the range of about $1\sim10$ times normal
nuclear density and temperature not too high $T=10\sim50$MeV, our
treatment gets the result that
\begin{enumerate}
\item Cluster mass $M_{cl}$ is in the range $500\sim1100$MeV and
increase with increasing density.
\item Quark chemical potential is in the range $200\sim 400$MeV
and increase with increasing density
\item Available volume is about $15\%\sim 30\%$ while $80\%\sim 90\%$
of baryon number is in clusters. In other words, most volume and
baryon number is in clusters at moderate densities similar to
Clark's two flavor system.
Above some temperature between
$30$MeV to $40$MeV cluster fraction decrease with increasing density while
below this temperature the behavior is the opposite.
\end{enumerate}
As mentioned above, the excluded volume treatment interaction
is taken into account only through influence of finite volume on
phase space integral which is a very rough approximation.
However, it suggests a possibility that
clusters can appear also in strange quark matter and
in which clusters takes most volume and baryon number. This
in turn favors the picture that clusters immersed in small amount of
free quarks which somehow resembles positive ion of normal metal immersed in
electron gas.

On the other hand, while positive ions in metal are considered
classical particles, in excluded volume method 3 quark clusters are
still too light to be treated classically. In the future we plan to
calculate clustering which involves more heavier species such as
H-dibaryon \cite{jaffe77} or even `quark alpha' \cite{michel88} with
18 quarks (six of each flavor). For heavy clusters of mass $m$ GeV
with baryon number $B$ we can simply compare the non-relativistic
expression of thermal wavelength
$\lambda=\sqrt{2\pi\hbar^2/(mk_BT)}$ to mean particle separation
$l=n^{-1/3}=(\nu/B)^{1/3}$ to work out a temperature scale above
which the wave packets of clusters no longer strongly overlap. It
can be show that this temperature is
\begin{equation}
\frac{T}{\mbox{MeV}}\simeq 72B^{-2/3}\frac{\mbox{GeV}}{m}\left(\frac{\nu}{\nu_0}\right)^{2/3}
\end{equation}which means that for clusters of mass that equals $6m_\Lambda\simeq
6\mbox{GeV}$ (ignoring a possible binding energy which would reduce
this mass) at density $3\sim 4\nu_0$ when temperature grows well
above $7\sim 9MeV$ cluster would behave like classical particle
which makes them capable of forming lattice.

\bigskip
I am grateful to pulsar group here at Peking University and Professor Efrain J. Ferrer
and Vivian de la Incera from University of Texas El Paso for inspiring discussions.

\def\Discussion{
\setlength{\parskip}{0.3cm}\setlength{\parindent}{0.0cm}
     \bigskip\bigskip      {\Large {\bf Discussion}} \bigskip}
\def\speaker#1{{\bf #1:}\ }
\def\endDiscussion{}

\Discussion

\speaker{Prof. Qi-Ren Zhang (Peking University)} Do you think hadrons are clusters, if not
what is the difference between hadrons and clusters? The second question is that there
are models for nuclear matter based on the quark crystal model or bag crystal model.
Is there any relations between your ideas and this model? This model was worked out
about 20 years ago and some are quite impressive results for example they can reproduce
nuclear data.

\speaker{Xuesen Na} For the first question, the hadronic vacuum have been move out of
the entire region which can be seen from equation of state we used.

\endDiscussion

\end{document}